\documentclass[%
 reprint,
 superscriptaddress,
bibnotes,
 amsmath,
 amssymb,
 aps,
 prl,
]{revtex4-2}

\usepackage{array}
\usepackage{tabularx}
\usepackage{multirow}
\usepackage{graphicx}
\usepackage{dcolumn}
\usepackage{bm}
\usepackage[colorlinks,
			linkcolor=blue,
            anchorcolor=blue,
            citecolor=blue,
            urlcolor=blue,
            breaklinks=true]{hyperref}

\newcommand{\eref}[1]{Eq.~(\ref{#1})}
\newcommand{\fref}[1]{Fig.~\ref{#1}}

\newcommand{\p}{\ensuremath{\partial}}

\newcommand{\df}{\ensuremath{\mathrm{d}}}
\newcommand{\bra}[1]{\ensuremath{\langle #1|}}
\newcommand{\ket}[1]{\ensuremath{|#1 \rangle}}

\newcommand{\ie}{\emph{i.e.}}
\newcommand{\eg}{\emph{e.g.}}

\newcommand{\cd}{\ensuremath{\mathrm{C}^\circ}}

\begin{document}

\title{Photon discerner: Adaptive quantum optical sensing near the shot noise limit}
\author{F. Bao}\email{baof@purdue.edu}
\author{L. Bauer}
\author{A. E. Rubio L\'opez}
\author{Z. Jacob}\email{zjacob@purdue.edu}
\affiliation{Birck Nanotechnology Center, School of Electrical and Computer Engineering, Purdue University, West Lafayette, IN 47907, USA}
\date{\today}

\begin{abstract}
    Photon statistics of an optical field can be used for quantum optical sensing in low light level scenarios free of bulky optical components. However, photon-number-resolving detection to unravel the photon statistics is challenging. Here, we propose a novel detection approach, that we call `photon discerning', which uses adaptive photon thresholding for photon statistical estimation without recording exact photon numbers. Our photon discerner is motivated by the field of neural networks where tunable thresholds have proven efficient for isolating optimal decision boundaries in machine learning tasks. The photon discerner maximizes Fisher information per photon by iteratively choosing the optimal threshold in real-time to approach the shot noise limit. Our proposed scheme of adaptive photon thresholding leads to unique remote-sensing applications of quantum DoLP (degree of linear polarization) camera and quantum LiDAR. We investigate optimal thresholds and show that the optimal photon threshold can be counter-intuitive (not equal to 1) even for weak signals (mean photon number much less than 1), due to the photon bunching effect. We also put forth a superconducting nanowire realization of the photon discerner which can be experimentally implemented in the near-term. We show that the adaptivity of our photon discerner enables it to beat realistic photon-number-resolving detectors with limited photon-number resolution. Our work suggests a new class of detectors for information-theory driven, compact, and learning-based quantum optical sensing.
\end{abstract}

\maketitle

\section{Introduction}
Parameter estimation of complex optical fields lies at the heart of classical and quantum optical sensing \cite{Yuan2023,Pirandola2018,Xia2023,Walmsley2015}, computing \cite{Bogaerts2020,Jeremy2007,Lvovsky2009}, and communication \cite{Yin2012,Pirandola2015,Ursin2007}. Traditional photodetectors aim at accurately measuring the optical field strength. An intensity detector, for example, accumulates input photons in a given integration time and responds only to the average strength of the optical field. Sensing of other field attributes, such as the wave vector, frequency, polarization and phase, is usually performed using bulky optical components like lens, gratings or prisms, polarizers, and interferometers \cite{rastogi2014phase,Tyo2006}.

Processing of quantum information brings up the necessity of counting the exact photon number, $n$, for example, to distinguish or prepare quantum optical states \cite{Namekata2010} and to implement linear-optics quantum computing \cite{Knill2001}. The photon counting statistics, $p(n)$, is a signature of light sources and depends on the mode composition of the optical field \cite{Knoll2023}. Quantum optical sensing, \eg, passive quantum remote sensing, through photon statistics \cite{Matthews2016} is thus a promising route to overcome the challenges of operating bulky optical components in low light level scenarios. An ideal photon-number-resolving detector (PNRD) that can register the photon statistics, however, is an outstanding goal that has only been approximately implemented for low photon numbers. For PNR schemes involving space-multiplexing \cite{Divochiy2008,Miatto2018} or beam-splitter cascades \cite{Pieter2001}, a strong restriction arises from the timing resolution, \ie, each of the sub-pixel or port and their readout must have very precise and predictable timing in order for the detector signals to be properly correlated. Those PNRDs are also impractical to scale up to a 2D array for imaging applications. For time-multiplexing \cite{Sperling2015,Avenhaus2010}, amplitude-multiplexing \cite{Kardynal2008} or the transition-edge \cite{Miller2003} PNRDs, the signal detection mechanism becomes complex as it is scaled. Either a higher-bit analog-to-digital conversion (ADC) is needed for resolving different amplitudes, or a larger detector bandwidth/faster ADC speed is needed for resolving different frequencies. Only recently has photon number resolution been scaled up to 100 photons \cite{Cheng2023}. Instead, single-photon detectors (SPD) that can only tell the presence/absence of photons are more widely used due to their relatively easy implementations \cite{Jahani2020,yang2020single}. SPDs lose the photon number information and usually cannot work at the shot noise limit of optimal quantum parameter estimation. A new class of sensors working near the shot noise limit without registering the exact photon number can bridge the gap and is urgently needed for quantum optical sensing.

With the fast development of artificial intelligence, there is a broad range of emerging applications where the measurement of accurate signal strength is not necessary. In classification problems, neural networks use nonlinear activation functions \cite{Rasamoelina2020} like sigmoid to extract the desired information. Sigmoid activations after trainable weights act like tunable thresholds, outputting binary value 1 when the signal is above the threshold while outputting 0 otherwise. They lose the information of accurate signal strength but are very efficient in classification \cite{Lin2017}.

\begin{figure}[tb]
    \centering
    \scalebox{1}{\includegraphics{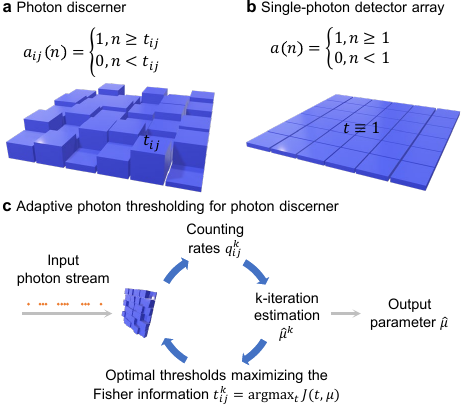}}
    \caption{Schematic of (a) the photon discerner (PD), in comparison with (b) an ideal single-photon detector (SPD) array, both with thresholded response functions to $n$ input photons, $a(n)$. The PD has a tunable and spatially-varying threshold, $t_{ij}$, at pixel position $(i,j)$. In contrast, the SPD array has a constant and uniform threshold, $t\equiv 1$. (c) Adaptive workflow of the PD. In the $k$-th iteration, a PD measures the counting rate $q^k_{ij}$ and gives the estimation $\hat{\mu}^k$ for the unknown parameter $\mu$. Based on the $k$-th estimation, PD updates the optimal threshold $t^k_{ij}$ to maximize the Fisher information $J(t,\mu)$ for adaptive quantum optical sensing near the shot noise limit.}
    \label{fig:schematic}
\end{figure}

Motivated by neural networks, here we propose a novel detection scheme of photon discerning (PD, see \fref{fig:schematic}), which uses spatially and temporally tunable photon thresholds for optimal threshold detection, approaching the performance of ideal photon-number-resolving detection.
We put forth two applications in quantum optical sensing which are enabled by our photon discerner. The first is a quantum camera which measures DoLP (degree of linear polarization) without a polarizer, and the second is a quantum LiDAR. We investigate optimal thresholds and show that the optimal threshold can be non-trivial (\ie, $t\neq 1$) even for weak signals with mean photon number $N\ll 1$, due to photon bunching. We further propose a superconducting nanowire realization of the photon discerner and show how it can beat many practical implementations of photon-number-resolving detectors.

\section{Photon discerner}
A binary threshold detector, or on/off detector, such as a single photon detector, is already widely adopted in industry and academia. Single-photon detectors are widely used as they have easier implementation than photon-number-resolving detectors. It has been appreciated that these on/off detectors can recover photon statistics without photon counting \cite{Guido2005,Harder2014,Han2018,Miatto2018,Brida06} and also reconstruct the Glauber $P$ or Wigner $W$ function for quantum optical states \cite{Alfredo2015,Alessia2009}.

Our proposed photon discerner (PD) introduces a new degree of freedom, \ie, the threshold adaptivity, to a binary detector. Note that tunability at the sensor level has recently been demonstrated to enable transformative learning-based optical sensing \cite{Yuan2023,Ma2022}. PD is thus a promising hardware implementation for a learning-based sensor. The uniqueness of PD is that the learnt adaptivity is guided by maximizing the Fisher information. This information optimization procedure can even occur through real-time learning if permitted by hardware reconfiguration speeds and edge computing resources. The response function of an ideal PD denoted by $a(n,t)$ is a step function at photon threshold $t$,
\begin{equation}
    a(n,t)=\Theta(n-t).
\end{equation}
It registers a logical count of 1 if the incident $n$-photon wavepacket has the number of photons larger than or equal to the threshold $t$, or a logical count of 0 otherwise, within a specific measurement window. The counting rate of PD is a function of the threshold $t$ and the photon statistics $p(n)$ of the input optical field,
\begin{equation}\label{eq:pdcr}
    q(t) = \sum_{n=0}^\infty a(n,t)p(n) = 1-\sum_{n=0}^{t-1}p(n).
\end{equation}
The counting rate gives the probability of `clicking' of PD in the given measurement window. The `no-click' probability is therefore given by $1-q(t)$. For one measurement window, the Fisher information of an unknown parameter $\mu$ by ideal photon-number-resolving detection (shot-noise limit) is given by the photon statistics $p(n)$,
\begin{equation}\label{eq:snl}
    J_0(\mu) = \sum_{n=0}^\infty\frac{[\p_\mu p(n)]^2}{p(n)}.
\end{equation}
In comparison, the Fisher information by a photon discerner is given by the counting rate $q(t)$ in the measurement window,
\begin{equation}\label{eq:FIPDDoLP}
    J(t,\mu) = \frac{[\p_\mu q(t)]^2}{q(t)} + \frac{[\p_\mu [1-q(t)]]^2}{1-q(t)} = \frac{[\p_\mu q(t)]^2}{q(t)[1-q(t)]}.
\end{equation}
PD loses photon number information, and rigorously one can prove that $J_0(\mu)\ge J(t,\mu)$ for all $t$. However, PD is adaptive. If the threshold $t$ of PD is set to the optimal photon threshold,
\begin{equation}
    t = \mathrm{argmax}_tJ(t,\mu),
\end{equation}
PD can maximize the Fisher information in the given counting window,
\begin{equation}\label{eq:FIPDmax}
    J(\mu)=\mathrm{max}_tJ(t,\mu).
\end{equation}
We find that adaptive PD outperforms single-photon detectors, \ie, $J(\mu)\ge J(1,\mu)$, and can approach ideal photon-number-resolving detectors, \ie, $J(\mu)\approx J_0(\mu)$, see the schematics in \fref{fig:schematic}(c). Particularly, for realistic PNRDs with limited photon number resolution, the summation in \eref{eq:snl} is truncated (see \eref{eq:M}) and the sum can drop below the Fisher information of a PD, which means PD can beat realistic PNRDs. The fact that PD can forget the photon number without losing much information can be seen from a basic example of discriminating Fock states $\ket{n-1}$ and $\ket{n}$, where photon thresholding at $t=n$ gives the optimal detection. We emphasize that the above argument also holds in general applications beyond state discrimination.

\section{Quantum DoLP camera}
High-precision polarimetric imaging is of fundamental importance in geoscience \cite{Tyo2006}, infrared sensing \cite{Gurton2014}, and astrophysics, for example, to characterize the cosmic magnetic fields \cite{Kulsrud2008,wang2023} and to discover exoplanets \cite{Berdyugina2011} and the existence of organic bio-molecules therein \cite{Jeremy1998}. Particularly, polarimetric imaging of the cosmic microwave background is crucial to test the standard cosmological model \cite{Kovac2002}. Though polarization can be measured down to the single-photon level for qubits in quantum computing \cite{Knill2001,Kok2007,Wang2019}, accurately measuring the degree of polarization of a photon stream is still challenging. State-of-the-art Cartesian polarizers have a finite extinction ratio, usually $\lesssim 1\times10^5$ for visible light, and $\lesssim 1\times10^3$ for the infrared. The inverse of the extinction ratio gives the uncertainty of the degree of linear polarization (DoLP). For $s$-$p$ polarizers, the finite spatial coherence (\ie, the divergence angle of the propagation direction) immediately sets the uncertainty of DoLP. Metasurface-based approach has been a new frontier in polarimetric imaging \cite{Noah2019} but has poor precision. Photoelastic modulation with Lock-in detection holds the record to measure DoLP down to around $1\times10^{-7}$ \cite{kemp1987optical,berdyugin2018high}. The ultimate difficulty in accurate metrology arises from the perturbation to the optical field of interest by the optical components themselves, including lens and polarizers. Furthermore, for deep ultra-violet with wavelength around tens to 100 nanometers, polarizers are not available. DoLP of X-rays or gamma-rays is instead measured through the photoelectric effect \cite{costa2001efficient}. We propose a quantum DoLP camera based on pinhole imaging and photon discerner, free of optical components, to overcome the above challenges.

\begin{figure*}[tb]
    \centering
    \scalebox{1}{\includegraphics{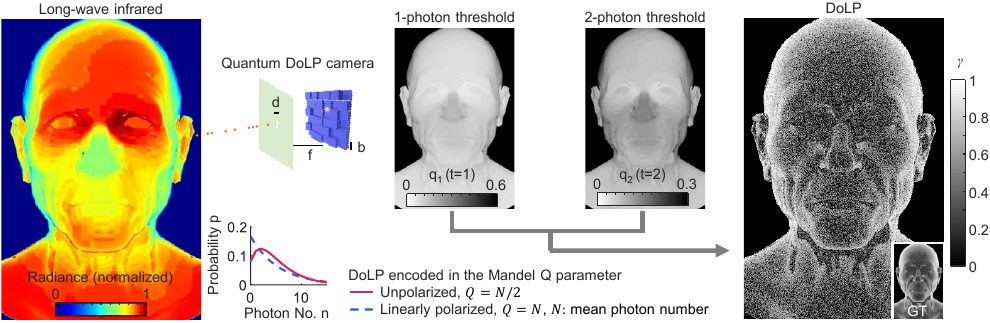}}
    \caption{Photon discerner enables the `quantum DoLP camera' that measures the degree of linear polarization without polarizers. In the long-wave infrared (LWIR) spectral range, a human face strongly emits thermal radiation. This renders the face textureless in traditional thermal imaging (left image), famously known as the `ghosting effect' \cite{Bao2023,Gurton2014}. In contrast, the quantum DoLP camera uses a pinhole camera and a photon discerner with adaptive photon thresholding, recording the face by logic counts with varying thresholds. From the logic count rates $q_{1(2)}$ for 1- and 2-photon thresholds, respectively, quantum DoLP camera directly derives the DoLP (right image) revealing otherwise hidden geometric facial features. Both mean photon number $N$ (proportional to radiance) and DoLP $\gamma$ had been normalized into the interval between 0 and 1 to ease computation in demonstrating the quantum DoLP camera. Ground truth DoLP (inset of the right image) of the face and photon streams were simulated by Monte Carlo path tracing and sampling, see Appendix~\ref{apd:mcpt}. DoLP: degree of linear polarization. GT: ground truth. Image size: $\mathrm{Height\times Width=480\times320}$.}
    \label{fig:dolp}
\end{figure*}
For a pinhole of diameter $d$ at a distance $f$ in front of a photon discerner, as illustrated in \fref{fig:dolp}, the coherence radius of the optical field at the sensor plane is given by $b_\mathrm{coh} = \lambda f/d$, and the coherence time of the optical field is $\tau_\mathrm{coh}=\lambda^2/c\Delta\lambda$, where $\lambda$ is the central wavelength of the optical field, $\Delta\lambda$ is the bandwidth, and $c$ is the speed of light. Let $b$ denote the discerner pixel size and $B$ the pixel pitch size (pixel to pixel distance). We require that one pixel focuses on one spatial and temporal mode, and that neighboring pixels focus on different spatial modes. This leads to $b \le b_\mathrm{coh} \le B$ and $\tau \le \tau_\mathrm{coh}$, $\tau$ being the measurement time. Explicitly, we have the following requirements for the quantum DoLP camera,
\begin{equation}\label{eq:dolpcon}
    \begin{split}
        \tau &\le \lambda^2/c\Delta\lambda,\\
        b &\le \lambda f/d \le B,\\
        d &\approx \sqrt{2\lambda f}.
    \end{split}
\end{equation}
Note that the last equation gives the optimal pinhole size in order to form a sharp image. Also note that the small pinhole size and the short measurement time likely lead to low photon numbers and multiple counting windows are needed.

Under the condition of \eref{eq:dolpcon}, the optical field to be detected by a discerner pixel is a superposition of two orthogonal polarization modes. Each polarization mode is a thermal state. The joint state of photons of two modes is given by
\begin{equation}
    \begin{split}
    \hat{\rho} &= \hat{\rho}_\mathrm{th}(N_{\parallel}) \otimes \hat{\rho}_\mathrm{th}(N_{\perp}),\\
    N_{\parallel} &= \frac{N}{2}(1+\gamma),\\
    N_\perp &= \frac{N}{2}(1-\gamma),
    \end{split}
\end{equation}
where $\hat{\rho}_\mathrm{th}(x)$ is the density operator for a thermal state of mean photon number $x$, $N$ is the total mean photon number, and $\gamma$ is the degree of linear polarization, $0\leq \gamma \leq 1$. The joint photon statistics, $p(n) = \bra{n} \hat{\rho} \ket{n}$, is a convolution of two Bose-Einstein distributions and is given by
\begin{equation}\label{eq:dolpps}
\begin{split}
    p(n) &= \sum_{k=0}^n \frac{N_\parallel^k}{(N_\parallel+1)^{k+1}}\frac{N_\perp^{n-k}}{(N_\perp+1)^{n-k+1}}\\
    &=\frac{1}{N_\parallel - N_\perp}\left[ \frac{N_\parallel^{n+1}}{(N_\parallel+1)^{n+1}} - \frac{N_\perp^{n+1}}{(N_\perp+1)^{n+1}} \right].
\end{split}
\end{equation}
It can be readily shown that when $\gamma=1$, the photon statistics remains a Bose-Einstein distribution, $p(n) = N^n/(N+1)^{n+1}$, and when $\gamma=0$, the photon statistics transits to a negative binomial distribution, $p(n) = (n+1)(N/2)^n/(N/2+1)^{n+2}$. Intuitively, for a given total intensity of the input light field, DoLP affects the profile of the joint photon statistics and hence is encoded in the Mandel Q parameter, see \fref{fig:dolp}.

We first show how DoLP can be measured by photon statistics with the non-adaptive mode of photon discerner since it is more intuitive. Explicitly, the parameters $N$ and $\gamma$ are in the first two terms of the photon statistics,
\begin{align}
    p(0) &= \frac{1}{(N/2+1)^2 - N^2\gamma^2/4},\\
    p(1) &= \frac{N(N/2+1) - N^2\gamma^2/2}{[(N/2+1)^2 - N^2\gamma^2/4]^2}.
\end{align}
They can be recovered by checking the counting rate of PD with threshold set at 1-photon and 2-photon,
\begin{gather}
    N = \frac{q_1+q_2-2q_1^2}{(1-q_1)^2},\\
    \gamma = \frac{\sqrt{(q_2+2-3q_1)^2-4(1-q_1)^3}}{q_1+q_2-2q_1^2},
\end{gather}
where $q_1\equiv q(1)$ and $q_2\equiv q(2)$. 

\fref{fig:dolp} illustrates the workflow of the non-adaptive mode of the quantum DoLP camera with a pertinent example in infrared thermal imaging. Long-wave infrared thermal imaging is important for night-vision enhancement in defense, facial recognition, and autonomous navigation \cite{Bao2023}. The challenge in thermal imaging is the loss of geometric textures which are crucial for object recognition and ranging. DoLP depends on the geometric orientations and hence carries geometric textures. \fref{fig:dolp} shows that a traditional thermal camera sees a face without textures. This phenomenon where long-wave infrared thermal images lose contrast/texture is addressed as ghosting. In contrast, the quantum DoLP camera capturing additional degrees of freedom through photon statistics, \ie, DoLP, recovers geometric facial features.

\begin{figure*}[tb]
    \centering
    \scalebox{1}{\includegraphics{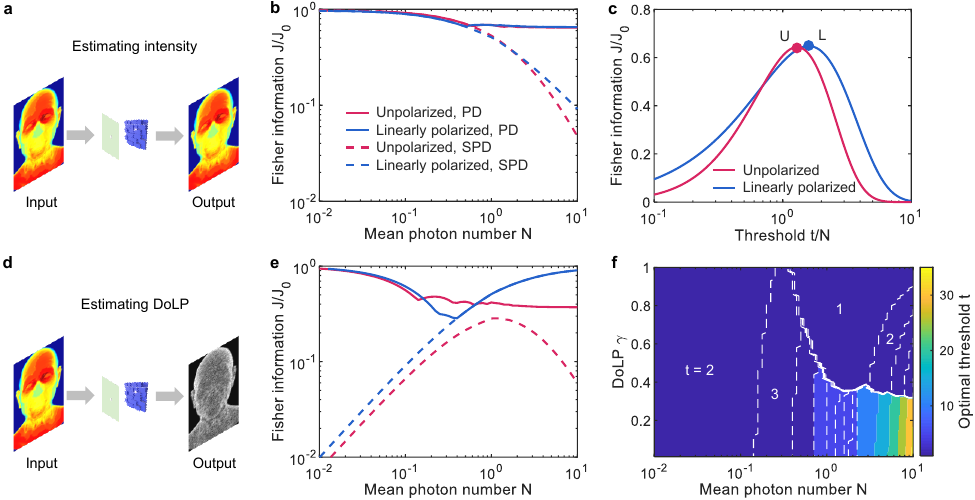}}
    \caption{Quantum DoLP camera based on photon discerner extracts information of optical features near the shot-noise limit with nontrivial thresholds. (a-c), Estimation of the mean photon number. (d-f), Estimation of the degree of linear polarization. (a) and (d) are schematics of the tasks. (c) The optimal threshold to estimate the mean photon number scales with the mean photon number itself. (f) However, for DoLP estimation, two-photon threshold is always optimal for weak signals ($N\le 0.1$). This nontrivial threshold results from the different coalescent photon statistics of unpolarized light, in contrast to linearly polarized light. $J$: Fisher information extracted by the photon discerner. $J_0$: the full amount of information in input photon streams that can be extracted by ideal photon-number-resolving detection, \ie, the shot-noise limit. PD: photon discerner. SPD: single-photon detector. In (c), the equation for unpolarized light is $y=8x^4/(2x+1)/(e^{2x}-2x-1)$. The equation for polarized light is $y=x^2/(e^x-1)$. U\,$\approx(1.29,0.64)$. L\,$\approx(1.59,0.65)$.}
    \label{fig:dolpopt}
\end{figure*}
Note that the above non-adaptive mode of the quantum DoLP camera can be implemented by a single-photon detector with varying quantum efficiencies. However, the adaptive mode of the quantum DoLP camera working at the optimal threshold maximizes the Fisher information per photon and hence has a higher photon efficiency. \fref{fig:dolpopt} shows the maximal Fisher information $J(\mu)=\mathrm{max}_tJ(t,\mu)$ that can be achieved by PD and its optimal threshold $t = \mathrm{argmax}_tJ(t,\mu)$. It can be seen in \fref{fig:dolpopt}(b and e) that PD (solid curves) estimates both unknown parameters near the shot noise limit (\ie, $J/J_0 \sim 1$), outperforming the single-photon detector (dashed curves). In estimating intensity, the optimal threshold $t$ is proportional to the mean photon number $N$, as can be seen in \fref{fig:dolpopt}(c) where the Fisher information reaches the maxima at fixed points U and L. For weak signals with $N\ll 1$, the optimal threshold is 1. Therefore, both PD and SPD reach the shot-noise limit in \fref{fig:dolpopt}(b) in the $N\to 0$ limit. It is expected that for strong signals with $N>1$, the optimal threshold is larger than 1, and hence PD outperforms SPD in the $N\to\infty$ limit in \fref{fig:dolpopt}(b). But counter-intuitively, the optimal threshold to estimate the degree of linear polarization has a non-trivial pattern. The optimal threshold remains $t=2$ even for extremely weak optical fields ($N\ll 1$) and cannot be decreased by applying optical attenuators. Therefore, PD outperforms SPD for all $N$ in \fref{fig:dolpopt}(e), for unpolarized light. This strongly indicates that photon discerners working at optimal thresholds have unique features that cannot be implemented by single-photon detectors.

\begin{figure}
    \centering
    \scalebox{1}{\includegraphics{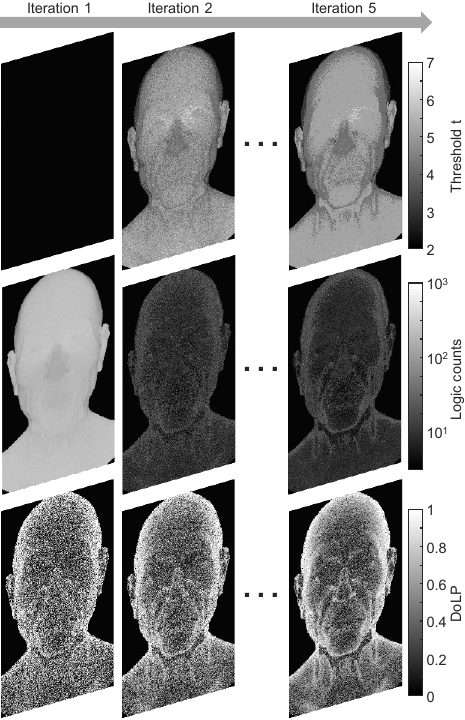}}
    \caption{Photon discerner as an adaptive sensor for adaptive imaging. As demonstrated in the case of quantum DoLP camera, photon discerner learns optimal thresholds from observed data and iteratively maximizes the Fisher information per photon. This adaptive imaging of DoLP achieves a given precision using nearly half amount of counting windows of the non-adaptive imaging with uniform 1- and 2-photon thresholds as shown in \fref{fig:dolp}. Used parameters are the same as \fref{fig:dolp}.}
    \label{fig:adaptive}
\end{figure}

Making use of optimal thresholds, \fref{fig:adaptive} shows the workflow of the adaptive quantum DoLP camera, according to the general adaptive strategy in \fref{fig:schematic}(c). In this illustration, photon discerner uses a pair of thresholds in each iteration, one optimal for estimating the mean photon number, and the other optimal for estimating the DoLP. The total counting windows are 1000 for each threshold and each iteration to estimate the counting rates. From the counting rates, we estimate the unknown parameters $\{N,\gamma\}$ according to \eref{eq:pdcr} and \eref{eq:dolpps}. We start with the threshold pair of $t=\{1,2\}$ in the initialization. After estimating the unknown parameters $\{N,\gamma\}$ in each iteration, we update the optimal threshold pair according to the results in \fref{fig:dolpopt}. The adaptive loop ends when a given precision criterion is met.
\fref{fig:adaptive} shows the optimal threshold pattern, logic counts (in 1000 counting windows), and estimation of DoLP in each iteration. This adaptive imaging of DoLP achieves the given DoLP precision using nearly half amount of counting windows of the non-adaptive imaging with uniform 1- and 2-photon thresholds as shown in \fref{fig:dolp}. The DoLP error (mean absolute difference) in the 5th-iteration is about $0.11$.

We emphasize that the precision of DoLP depends only on the number of counting windows, $N_\mathrm{c}$. By increasing $N_\mathrm{c}$, the total Fisher information can be accumulated to $N_\mathrm{c}J$, with $J$ given in \eref{eq:FIPDDoLP}. Therefore, we have $\delta\gamma\propto 1/\sqrt{N_\mathrm{c}}$, which means the precision of DoLP can go down below $10^{-7}$ with the number of counting windows over the order of $10^{14}$. For a counting window at the timescale of a nanosecond, this corresponds to a day of measurement in total.

\section{Quantum LiDAR}
Now we illustrate the photon discerner for the detection of a coherent signal in the thermal background. We show the existence of non-trivial thresholds and their advantages over traditional single-photon detectors. This can lead to quantum LiDAR and quantum radar with improved detectivity for self-driving cars and robotics, and it is also important for detecting interstellar maser signal in the cosmic microwave background \cite{wenner2022long}.

Without loss of generality, we use a polarizer in front of the quantum LiDAR to filter the thermal noise into a single mode, for the proof of principle. The input optical field in this case is a displaced thermal state,
\begin{equation}\label{eq:dthrho}
    \hat{\rho} = D(\alpha)\cdot \hat{\rho}_\mathrm{th}[N(1-g)]\cdot 
 D^\dagger(\alpha),
\end{equation}
where $D$ is the displacement operator, $|\alpha|^2 = Ng$, and $g$ is a classical parameter depicting the intensity fraction of the coherent signal in the thermal background, $0\le g \le 1$.

\begin{figure*}
    \centering
    \scalebox{1}{\includegraphics{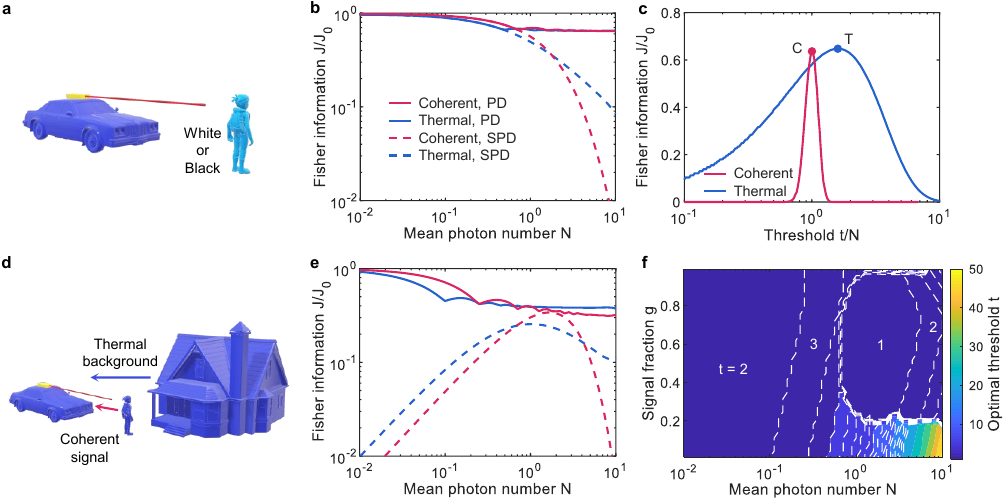}}
    \caption{Quantum LiDAR based on photon discerner extracts information of optical features near the shot-noise limit with nontrivial thresholds. (a-c), Estimating the mean photon number. (d-f), Estimation the signal fraction. (a) and (d) are schematics of the tasks. (c) The optimal threshold for the mean photon number scales with the mean photon number itself. (f) However, in estimating the signal fraction, two-photon threshold is always optimal for weak signals $N\le 0.1$. This nontrivial threshold results from the photon bunching effect of thermal light sources, in contrast to coherent light sources. $J$: Fisher information extracted by photon discerner (PD). $J_0$: the shot-noise limit. SPD: single-photon detector. C\,$=(1,2/\pi)$. T\,$\approx (1.59,0.65)$. $g$: the fraction of coherent signal in thermal background.}
    \label{fig:lidaropt}
\end{figure*}

Note that a thresholded quantum LiDAR has been reported in the literature \cite{Cohen2019}, where the strength of the thermal background is assumed to be known and the metric of signal-to-noise ratio (SNR) has been discussed. The key difference of our work with respect to the above reference is that we do not assume a known thermal background, which is usually the case in realistic applications, and we use the Fisher information as the proper metric. We emphasize that improving SNR will result in a misleading conclusion that the infinitely high threshold is the best as it gives the highest SNR. However, the higher the threshold, the more signal is lost. Note that an infinitely high threshold technically loses all the useful information, even though the noise is completely suppressed. In stark contrast, maximizing the Fisher information, as can be seen in \fref{fig:lidaropt}, will yield a finite optimal threshold.

The photon statistics of the displaced thermal state given in \eref{eq:dthrho} reads
\begin{equation}
    \begin{split}
        p(n) = \bra{n}\hat{\rho}\ket{n} = \frac{1}{N(1-g)}\frac{e^{-g/(1-g)}}{n!}\times \\
        \int_0^\infty I_0[\frac{2\sqrt{xNg}}{N(1-g)}] \cdot e^{-x[1+\frac{1}{N(1-g)}]}\cdot x^{n} \df x.
    \end{split}
\end{equation}
Here, $I_0$ is the first-kind modified Bessel function of order 0. $N$ is the mean photon number, and $n$ is the photon number. When the fraction parameter $g\to 0(1)$, the above photon statistics reduces to a Bose-Einstein (Poisson) distribution. Similar to the case of quantum DoLP camera, quantum LiDAR based on photon discerner also exhibits non-trivial optimal thresholds, see \fref{fig:lidaropt}(f). For weak signals, to distinguish a coherent signal with respect to the thermal noise, the optimal threshold is 2 photons, instead of 1 photon.

\begin{figure*}
    \centering
    \scalebox{1}{\includegraphics{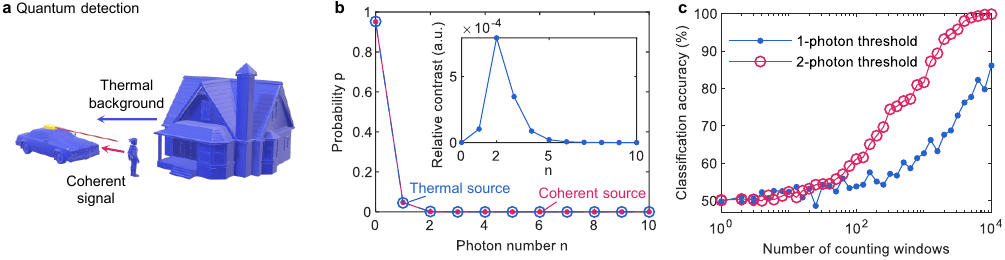}}
    \caption{Monte Carlo experiments demonstrating the optimal 2-photon threshold against the 1-photon threshold for quantum detection of the presence of coherent signals. (a) Schematics of the task. (b) Photon statistics show that for a weak optical field (mean photon number $N=0.1$), thermal and coherent light sources become almost identical. Inset: the highest contrast of their photon statistics locates at 2-photon bouching. (c) In classifying coherent signal and thermal noise of a given strength ($N=0.1$), quantum LiDAR based on a 2-photon-threshold photon discerner gives higher detection accuracy than a traditional LiDAR based on the single-photon detector.}
    \label{fig:lidar2}
\end{figure*}

We further test the non-trivial 2-photon threshold for quantum LiDAR. \fref{fig:lidar2} shows a Monte Carlo experiment in classifying the coherent laser signal against the thermal background noise, whose total intensity has been measured by a traditional photodetector ($N=0.1$).
We use $t=2$ for the quantum LiDAR based on a photon discerner and $t=1$ for the traditional LiDAR based on a single-photon detector. According to the Poisson (Bose-Einstein) distribution for the coherent (thermal) light, we know coherent signal with $N=0.1$ gives a higher counting rate than thermal noise with $N=0.1$, for the traditional LiDAR. In contrast, thermal noise gives a higher counting rate than the coherent signal for the quantum LiDAR. That is, $q_\mathrm{coh}(1)>q_\mathrm{th}(1)$, but $q_\mathrm{coh}(2)<q_\mathrm{th}(2)$. For a given coherent signal and a given thermal noise of strength $N=0.1$, we generate random photon streams according to their photon statistics. And for given numbers of counting windows, we derive the counting rates for both light sources and both LiDARs. We make classifications according to the above knowledge, \ie, traditional LiDAR picks the one with a higher counting rate as the `coherent' signal, while quantum LiDAR picks the one with a lower counting rate as the `coherent' signal. We repeat the experiment by 1000 times to derive the statistical classification accuracy. \fref{fig:lidar2}(c) clearly confirms that quantum LiDAR based on the photon discerner has an improved detection accuracy against the traditional LiDAR based on a single-photon detector.

\section{Superconducting nanowire realization of photon discerner}
\begin{figure}
    \centering
    \scalebox{1}{\includegraphics{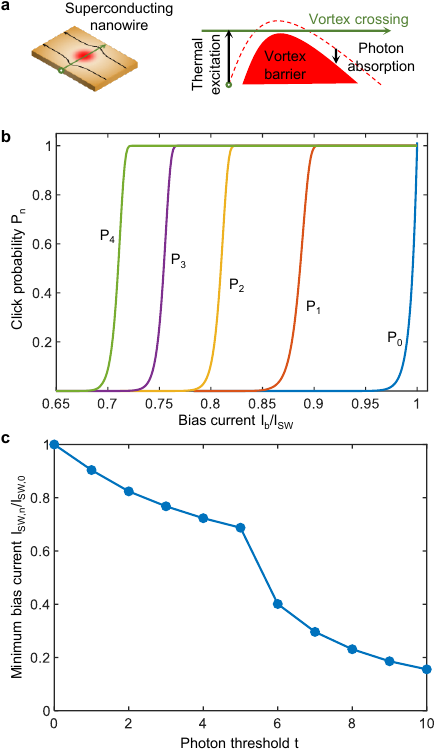}}
    \caption{Superconducting nanowire realization of the photon discerner. (a) Vortex crossing mechanism of the superconducting nanowire. (b) The click probability of the superconducting nanowire, $P_n$, for $n$ input photons as functions of the normalized bias current. (c) The minimum bias current to be applied on the superconducting nanowire for different photon thresholds $t$. $I_\mathrm{b}:$ Bias current. $I_\mathrm{SW,n}:$ Minimum bias current at which an $n$-photon event will cause a detector click with $P\ge 0.999$. $I_\mathrm{SW}=I_\mathrm{SW,0}$. The nanowire width, length, and thickness are $500\,\mathrm{nm}$, $500\,\mathrm{nm}$, and $50\,\mathrm{nm}$, respectively. WSi is used as the material of the nanowire, and the working temperature is $0.3\,\mathrm{K}$. The photon energy is $0.3\,e\mathrm{V}$, which corresponds to a wavelength around $4.1\,\mu\mathrm{m}$.}
    \label{fig:sn}
\end{figure}

Now we provide a superconducting nanowire \cite{Chiles2022,Esmaeil2021} realization of the photon discerner, following the framework of Ref.~\cite{Jahani2020}. The photon discerner is composed of a thin superconducting nanowire slightly biased below the superconducting critical current. Absorption of photons creates hot electrons. The presence of these hot electrons tends to lower the vortex barrier, allowing for easier vortex crossing, see the schematics in \fref{fig:sn}(a). Absorption of $n$ photons has a probability, $P_n$, to trigger a vortex crossing of the superconducting nanowire and eventually a click of the sensor. The click probability is dependent of the bias current that is tunable to implement photon thresholds. See Appendix~\ref{apd:sn} for the simulation details of our PD.

\fref{fig:sn}(b) shows the click probability of the superconducting nanowire for $n$ input photons. When the normalized bias current is around 0.9, the superconducting nanowire has a near-unity probability to click for $n\ge 1$ and a near-zero probability to click for $n=0$. As the normalized bias current decreases to around 0.82, the superconducting nanowire has a near-unity probability to click for $n\ge 2$ and a near-zero probability to click for $n<2$. When the normalized bias current further decreases, the superconducting nanowire acts exactly as the photon discerner with higher and higher photon thresholds. \fref{fig:sn}(c) shows the corresponding bias currents need to be applied for different photon thresholds.

For existing photon-number-resolving detectors, readout complexities like timing resolution and analog-digital conversion significantly restrict it from scaling up. The superconducting nanowire realization of the photon discerner places the sensor complexity in biasing rather than readout. Regardless of the input photon number, only a single signal of a single amplitude level needs to be timed. Additionally, the detector threshold is directly encoded in the detector response and can be modified without changing the hardware structure. Therefore, the superconducting nanowire realization of the photon discerner has a considerably simpler readout than existing PNRD. From \fref{fig:sn}(b), it can be seen that for large thresholds and low bias currents, the threshold function may not be an ideal step function. Appendix~\ref{apd:ft} shows the analysis of imperfect thresholds and the comparison of our photon discerner with respect to traditional intensity detectors. Also worth noting is that superconducting nanowire detectors can be made polarization independent \cite{Meng2022}, which is important to polarimetric imaging.

\section{Beating realistic PNRD}
We show how the adaptivity of a photon discerner enables it to beat realistic photon-number-resolving detectors. For a fair comparison across different sensors, we note that space-multiplexing PNRDs lose spatial resolution compared with the photon discerner, while time-multiplexing PNRDs lose temporal resolution. Therefore, we compare the photon discerner with the PNRD based on the transition-edge sensor model for PNRD. The conclusion is, however, general. The comparison is made for one counting window, \ie, the same measurement time or photon budget.

\begin{figure}
    \centering
    \scalebox{1}{\includegraphics{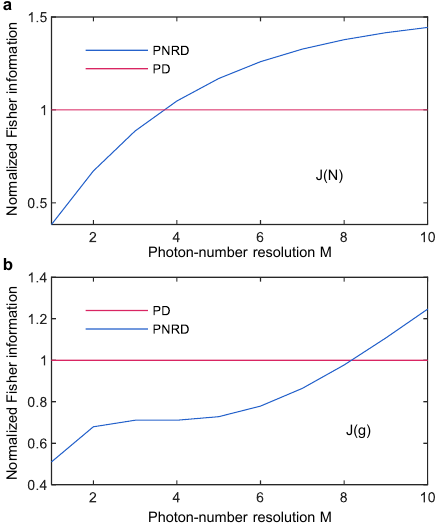}}
    \caption{Photon discerner beats a realistic photon-number-resolving detector for low photon-number resolution $M$. (a) Fisher information with respect to the mean photon number $N$, and (b) Fisher information with respect to the signal fraction $g$, for a PNRD and a PD as functions of the maximum number of resolvable photons $M$. In both (a) and (b), the Fisher information curves of PNRD are normalized by that of PD. PD beats PNRD in the low-$M$ region. Simulation is done for the quantum LiDAR applications with $N=3$ and $g=0.01$.}
    \label{fig:cmp2pnrd}
\end{figure}

We note the response function of a realistic PNRD to $n$ input photons can be described as
\begin{equation}
    a(n,M) = \min(n,M),
\end{equation}
where $M$ is the maximum resolvable number of photons. The key difference between a PD and a realistic PNRD is that PNRD is accumulative while PD is adaptive. PNRD always starts counting from 1, 2, 3 and so forth, but PD can directly set to a threshold of, \eg, 4-photon, without resolving lower photon numbers. This adaptivity of PD is beneficial when $n$-photon bunching carries more information than lower photon events. This can be most clearly seen in the aforementioned task of discriminating the Fock state $\ket{n}$ with $\ket{n-1}$, but it also holds true for general remote sensing applications as discussed below. \fref{fig:cmp2pnrd} shows the performance comparison between a PNRD and a PD, for the quantum LiDAR application.
The Fisher information for a PNRD is given by
\begin{equation}\label{eq:M}
    J(\mu) = \sum_{n=0}^{M-1}\frac{[\p_\mu p(n)]^2}{p(n)} + \frac{[\sum_{n=M}^\infty\p_\mu p(n)]^2}{\sum_{n=M}^\infty p(n)}.
\end{equation}
And the Fisher information for a PD is given in \eref{eq:FIPDmax}. In this case, the optimal threshold for estimating $N$ is calculated to be $t=6$ photons, and the optimal threshold for estimating $g$ is $t=18$ photons. 
For most PNRDs existing nowadays, they can only resolve below $M=10$ photons. When $M\ge t$, PNRD will have more photon number information than the PD. However, when $M<t$, PD has the potential to beat PNRDs.

In the low-$M$ region of \fref{fig:cmp2pnrd}, it can be seen that a PD captures more Fisher information than a realistic PNRD, for both parameters. In the limit of $M\to\infty$, a realistic PNRD approaches an ideal PNRD. A PD loses the information of photon number and hence cannot beat the ideal PNRD which sets the shot-noise limit. Note that the recent PNRD that can resolve 100 photons \cite{Cheng2023} is based on the superconducting nanowire. We believe its performance will further improve if our proposed threshold adaptivity is incorporated. Also note that a PD with counting rates for all thresholds can recover the photon statistics through $p(n)=q(n)-q(n+1)$.

\section{Conclusion}
We proposed a novel detection scheme of `photon discerning', with adaptive photon thresholding, to maximize the Fisher information per photon. Photon discerner enables novel applications of quantum DoLP camera and quantum LiDAR that beat traditional counterparts, approaching the shot noise limit without recording the exact photon numbers. We demonstrated the existence of non-trivial thresholds for both discriminating quantum optical states and general remote sensing applications. We also provided a superconducting nanowire realization of the photon discerner and shown its advantage over realistic photon-number-resolving detectors. Our work can lead to a new class of sensors for informative and compact quantum optical sensing, of importance to both fundamental science and practical applications.

\appendix
\section*{Appendix}
\section{Monte Carlo path tracing and photon streams}\label{apd:mcpt}
To simulate the thermal polarimetric imaging of the human face, it is common to model the face as a blackbody covered by a beam splitter. We use the Monte Carlo path tracing to render \fref{fig:dolp}, according to the following rendering equation,
\begin{equation}
    \Vec{S} = \hat{R}\times \Vec{S} + \hat{T}\times \Vec{E},
\end{equation}
where $\Vec{S} = (S_0,S_1,S_2)^\dagger$ is the Stokes vector of the thermal light field to describe its polarization state. Note that we have ignored the circular polarization for natural thermal light which is usually near zero. Here, the reflection matrix $\hat{R}$ and the transmittance matrix $\hat{T}$ are given by
\begin{equation}
    \hat{R} = 
    \begin{bmatrix}
       \frac{R_s+R_p}{2}  & \frac{R_s-R_p}{2} & 0 \\
        \frac{R_s-R_p}{2} & \frac{R_s+R_p}{2} & 0 \\
        0 & 0 & r_s\cdot r_p
    \end{bmatrix},
\end{equation}
and
\begin{equation}
    \hat{T} = 
    \begin{bmatrix}
       1-\frac{R_s+R_p}{2}  & \frac{R_s-R_p}{2} & 0 \\
        \frac{R_s-R_p}{2} & 1-\frac{R_s+R_p}{2} & 0 \\
        0 & 0 & \sqrt{(1-R_s)(1-R_p)}
    \end{bmatrix},
\end{equation}
where $r_{s(p)}$ is the amplitude reflectance for $s(p)$ waves given by the well-known Fresnel equations, and $R_{s(p)} = |r_{s(p)}|^2$ is the power reflectance for $s(p)$ waves. $\Vec{E} = E_0\times(1,\cos\theta,\sin\theta)^\dagger$ is the Stokes vector of self thermal emissions inside the face, $E_0$ is the blackbody radiation given by Planck's law, and $\theta$ is the polarization angle of the self emission with respect to the local face surface normal. Since thermal radiation is random, $\theta$ is an arbitrary value between 0 and $2\pi$ that will average out in Monte Carlo simulations. Without loss of generality, the refraction index of the face side of the beam splitter was assumed to be $1.1$ to compute the Fresnel coefficients, and the refraction index of the sensor side was set as 1. The face was at $37\,\cd$ and the environment was set as a uniform blackbody at $0\,\cd$. The final path tracing result gives the ground truth thermal radiance, $S_0$, and the ground truth DoLP, $\gamma = \sqrt{S_1^2+S_2^2}/S_0$. $S_0$ was normalized into the interval between 0 and 1, as the input mean photon number $N$ of \fref{fig:dolp}.

After getting the ground truth mean photon number $N$ and ground truth DoLP $\gamma$ from path tracing, we generate random photon streams for each face pixel according to the photon statistics in \eref{eq:dolpps}.

\section{Superconducting Nanowire Simulations}\label{apd:sn}
The superconducting nanowire simulations follow numerical modeling procedures detailed in \cite{Jahani2020}. We find click probabilities $P_n$ by assuming $n$ photons generate $n$ hot electrons at the same location. The hot electrons then diffuse out into the superconducting nanowire causing Cooper pairs to break into quasiparticles and form a hot-spot. This entire process is dictated by a series of diffusion equations dictating the evolution of the hot-electron and quasi-particle distributions. The hot-electron diffusion is given by
\begin{equation}
    \frac{\partial C_e(\Vec{r},t)}{\partial t} = D_e \nabla^2 C_e(\Vec{r},t)
\end{equation}
where $C_e$ is the hot-electron distribution density, and $D_e$ is the hot-electron diffusion coefficient. In this section, $t$ means time. Quasiparticle diffusion is given by
\begin{multline}
         \frac{\partial C_{qp}(\Vec{r},t)}{\partial t} =  D_{qp} \nabla^2  C_{qp}(\Vec{r},t) - \frac{C_{qp}(\Vec{r},t)}{\tau_r} + \\
     \frac{\varsigma h\nu}{\Delta\tau_{qp}}\left(\frac{n_{se,0}-C_{qp}(\Vec{r},t)}{n_{se,0}}\right)e^{-t/\tau_{qp}}C_e(\Vec{r},t)
\end{multline}
where $C_{qp}(\Vec{r},t)$ is the quasiparticle distribution density, $D_{qp}$ is the quasiparticle diffusion coefficient, $\varsigma$ is the quasiparticle conversion efficiency, $\tau_r$ is the recombination time, $\tau_{qp}$ is the quasiparticle multiplication process lifetime, $n_{se,0}$ is the density of superconducting electrons before absorption, $\Delta$ is the superconducting bandgap, and $h\nu$ is the energy of a photon with frequency $\nu$.
This hot-spot formation causes the nanowire current density $\Vec{j}(\Vec{r},t)$ to redistribute allowing for suppression of the vortex barrier. The current density is given by
\begin{equation}
    \Vec{j}(\Vec{r},t) = \frac{\hbar}{m} n_{se}(\Vec{r},t)\nabla\varphi(\Vec{r},t)
\end{equation}
where $\hbar=h/2\pi$ is the reduced Planck's constant, $m$ is the electron mass, $n_{se}(\Vec{r},t)$ is the density of superconducting electrons after absorption, $\varphi(\Vec{r},t)$ is the phase of the superconducting order parameter. Magnetic flux vortices located along the superconducting nanowire have some small probability of crossing this vortex barrier due to thermal energy in the system. The size of the vortex barrier $U_{\nu}(x_\nu,t)$ for different vortex positions $x_\nu$ is given by

\begin{multline}
    \frac{U_{\nu}(x_\nu,t)}{\varepsilon_0} =\frac{\pi}{W} \int_{\frac{\xi-W}{2}}^{x_\nu} \frac{n_{se}(x',t)}{n_{se,0}}tan\left(\frac{\pi x'}{W}\right)dx' - \\ \frac{2W}{I_{c,\nu}exp(1)\xi}\int_{-\frac{W}{2}}^{x_{\nu}}\frac{n_{se}(x',t)}{n_{se,0}} j_y(x',t)dx' 
\end{multline}
where $\varepsilon_0$ is the characteristic vortex energy, $W$ is the width of the nanowire, $\xi$ is the vortex core radius, and $I_{c,\nu}$ is the vortex critical current. As the vortex barrier becomes more suppressed, the vortex crossing rate $\Gamma_{\nu}(t)$ increases, thereby increasing the probability of vortex crossing $P_n(t)$. The crossing rate is given by
\begin{equation}
    \Gamma_{\nu}(t) = \alpha_\nu I_b exp\left(-U_{\nu,max}(t)/k_BT\right)
\end{equation}
where $\alpha_\nu$ is an experimentally measured fitting parameter, $I_b$ is the bias current, $U_{\nu,max}$ is the maximum potential barrier, $k_B$ is Boltzmann's constant, and $T$ is the device temperature. The probability of an $n$-photon click is given by
\begin{equation}
    P_n(t)=1-exp\left[-\int_0^t\Gamma_{\nu}(t')dt'\right]
\end{equation}
This entire process is drastically affected by bias current $I_b$. Therefore, the bias current can be used to modify the click probability for different incident photon numbers $n$ as demonstrated in Fig. \ref{fig:sn}.

\section{Flux-threshold detector}\label{apd:ft}
\begin{figure}
    \centering
    \scalebox{1}{\includegraphics{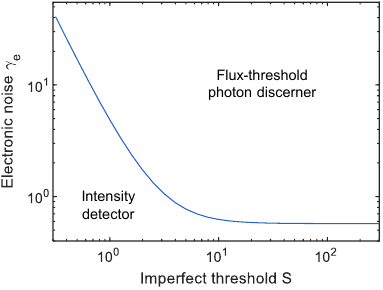}}
    \caption{Trade space of the imperfect flux-threshold photon discerner with intensity detectors. For a given sigmoid activation/response function characterized by $S$, the threshold-equivalent electronic noise defines the boundary curve of the trade space, above which the photon discerner outperforms intensity detectors.}
    \label{fig:flux}
\end{figure}

We analyze and show the trade space where photon discerners with imperfect flux thresholds can beat traditional intensity detectors. The imperfect threshold can be described by a parameter $S$ in the sigmoid response function,
\begin{equation}
    a(n,t) = \frac{1}{1+e^{-S(n-t)/\sqrt{t}}},
\end{equation}
where $t$ is the photon-flux threshold and $n$ is the photon number. When $S\to\infty$, the above response function reduces to a step function and the threshold becomes ideal. The counting rate of a PD now becomes
\begin{equation}\label{eq:fluxq}
    q(t) = \sum_{n=0}^\infty a(n,t)p(n).
\end{equation}
We define the threshold-equivalent electronic noise,
\begin{equation}
    \gamma_e = \frac{J_0}{J} - 1,
\end{equation}
according to \eref{eq:snl}, \eref{eq:FIPDDoLP} and \eref{eq:fluxq}. The threshold-equivalent electronic noise gives the corresponding electronic noise-equivalent power (NEP, in unit of photon number) of an intensity detector (normalized by photonic shot noise power) that can capture the same amount of Fisher information as a PD. That is, $\gamma_e = \mathrm{NEP}/N$, where $N$ is the photon number corresponding to the intensity of the optical field. \fref{fig:flux} shows the threshold-equivalent electronic noise curve, simulated for a sample input field of a coherent signal of $N=1000$. The Fisher information about parameter $N$ is evaluated at the threshold $t=N$. Above the threshold-equivalent electronic noise curve, a photon discerner outperforms intensity detectors by capturing more information.

\bibliography{PD}

\end{document}